\documentclass[aps,pra,twocolumn,showpacs,superscriptaddress]{revtex4}
\usepackage{epsfig}
\usepackage{bm,amsmath,amssymb,latexsym,mathrsfs,graphicx,enumerate}
\usepackage[mathcal]{euscript}
\usepackage{hyperref}

\newcommand{\be}{\begin{equation}}
\newcommand{\ee}{\end{equation}}

\begin{document}

%\title{The split-ring Josephson resonator as an artificial atom}
\title{Electrodynamics of a split-ring Josephson resonator in a microwave line}

\author{J.-G.~Caputo$^{~1}$, I. Gabitov$^{~2}$  and A.I.~Maimistov$^{~3,4}$}

\affiliation{\normalsize \noindent
$^1$: Laboratoire de Math\'ematiques, INSA de Rouen, \\
BP 8, Avenue de l'Universite,
Saint-Etienne du Rouvray, 76801 France \\
E-mail: caputo@insa-rouen.fr \\
$^2$: Department of Mathematics,\\
University of Arizona, Tucson, AZ, 85704, USA \\
E-mail: gabitov@math.arizona.edu\\
$^3$: Department of Solid State Physics and Nanostructures, \\
Moscow Engineering Physics Institute, \\
Kashirskoe sh. 31, Moscow, 115409 Russia \\
$^4$: Department of Physics and Technology of Nanostructures, \\
% REC Bionanophysics,\\
Moscow Institute for Physics and Technology, \\
Institutskii lane 9, Dolgoprudny, Moscow region, 141700 Russia \\
E-mail: amaimistov@gmail.com }

\date{\today }

\begin{abstract}
We consider the coupling of an electromagnetic wave to a split-ring Josephson
oscillator or radio-frequency SQUID in the hysteretic regime. This device
is similar to an atomic system in that it has a number of steady states.
We show that one can switch between these with a suitable short external
microwave pulse. The steady states can be characterized by their resonant lines
which are of the Fano type. Using a static magnetic field, we can shift these
spectral lines and lift their degeneracy.
\end{abstract}

\maketitle

\section{Introduction}

In the past decade artificial materials (metamaterials) have been
developed by inserting metal split-rings or rods as metaatoms into
dielectric materials. This way negative index materials have been
fabricated~\cite{Veselago:06}, \cite{Pendry:04}. These metaatoms
spaced periodically enable the medium to have an engineered
electromagnetic response. In principle such an artificial atom has
many advantages over a real atom. In real materials, the dipole
moment of the atom is very small and {the interaction with
electromagnetic field is small}. Therefore to realize significant
interactions, large densities are necessary. On the contrary an
artificial atom can be constructed with a high dielectric or a high
magnetic dipole moment so that it will respond strongly to
electromagnetic waves.

Many candidates for artificial atoms have been proposed, most of
them with intrinsic nonlinearities. Among them are resonant
conductive elements with inserted (nonlinear) diodes
\cite{Lapin:2003}, split-ring resonators loaded with varactor diodes
\cite{Katko:2010}, Kerr materials \cite{Zharov:2003} or laser
amplifiers \cite{Gabitov_Kennedy:2010}. Unfortunately many of these systems
exhibit large losses. To remedy this problem it has been
been suggested to work with superconducting split-ring resonator arrays.
An example is the study \cite{Chen:Yang:10} using
from high-temperature superconducting films. See also
the review \cite{Anlage:11} for more details.

Among these superconducting meta-materials are the split-ring Josephson
resonators {subsequently referred to as SRR-JJ} where a
Josephson junction, a superconducting weak link, is embedded in the slit.
In the Josephson community, this device is called an RF SQUID for
Radio Frequency Superconducting Quantum
Interference Device \cite{Barone} \cite{Likharev}.

The device we consider is shown in the left panel of Fig. \ref{f1}.
It is a split ring resonator in which is embedded a Josephson
junction. Practically it can be made using a ring like strip of
superconducting material where a small region was oxidized to make
the junction. The right panel of Fig. \ref{f1} shows the electric
representation of the device, an inductance $L$ for the strip and
the Resistive Shunted Junction (RSJ) model for the Josephson
junction (JJ). The latter represents the junction as a resistor $R$,
a capacity $C$ and the nonlinear element in parallel. This last
element is the sine coupling $I_c \sin (\Phi/ \Phi_0)$ where $\Phi$
is the magnetic flux and $\Phi_0$ is the reduced flux quantum $
\Phi_0 = \hbar/(2e)$.

The use of these devices in metamaterials was advocated by Lazarides
\cite{Lazarides:06} \cite{lt07} and by the authors
\cite{Maim:Gabi:10}. The electrodynamics of such metamaterials where
split-rings are coupled inductively was investigated in
\cite{Lazarides:06,lt07}. It leads to a 2D discrete sine-Gordon
equation. For weak external radio-frequency forcing and small
densities, this dilute gas of artificial atoms demonstrates a
nonlinear magnetic response \cite{Maim:Gabi:07}.

%In standard electronics the conjugate variables are voltages and
%currents while in superconducting electronics they are the fluxes
%and charges where a flux is defined as the time integral of a
%voltage. This is why we present the derivation in detail here. The
%device is assumed to be operating at low temperature so that losses
%are minimal and Josephson relations hold.

Such arrays of SRR-JJ are difficult to analyze because
the coupling between the artificial atoms is partly inductive,
partly capacitive. Another point is that the field couples to
the phase of the Josephson junctions so that the system should have
two components. For these reasons, in this article we study the
dilute system where each SRR-JJ can be considered as isolated
although coupled to an electromagnetic wave.
We consider the so-called hysteretic regime where the system has
controlled metastable states and show that one can switch from the
ground state to one of these excited states by applying a suitable
flux pulse.
Then we can detect these states by examining the
reflection coefficient of an electromagnetic wave incident on the
device, this is a kind of spectroscopy. More precisely, the resonance
observed are asymetric, of the Fano type. This is interesting
because it is observed for many plasmonic systems\cite{kivshar_fano}.
 
The article is organized as such. In section 2 we present the basic
equations of the model and analyze it's steady states. In section 3 we
show how to switch from one state to another by applying a suitable
flux pulse. Section IV discusses the scattering of an electromagnetic
wave by a SRR-JJ and this leads to a spectroscopy of the steady states.

\section{The model}

We consider here that the SRR-JJ is subject to irradiation by a
microwave field.
The equations describing the system light-ring are the
generalized pendulum equation for the flux
%(\ref{usrr_jj})
and the Maxwell equation for the electromagnetic field.
The Maxwell equations can be written as
\begin{align*}
\nabla\times\mathbf{E} &  =-\mathbf{B}_{,t},\qquad\nabla\cdot\mathbf{D}=0,\\
\nabla\times\mathbf{B} &
=\mathbf{E}_{,t},\qquad\nabla\cdot\mathbf{B}=0.
\end{align*}
where ${\bf E}$ and ${\bf H}$ are respectively the electric field and
the magnetic field,
and the subscritpts indicate the partial derivative.
The fields $ \mathbf{B}, {\bf H}$ and $\mathbf{E}, \mathbf{D}$ are
related by
\[
\mathbf{B}=\mu_{0}\mathbf{H}+\mathbf{M},\qquad\mathbf{D}=\varepsilon
_{0}\mathbf{E}+\mathbf{P},
\]
where $\mathbf{M}$ and $\mathbf{P}$ are respectively the magnetization
and polarisation of the medium.
In the dispersionless limit we can
write $\mathbf{D}=\varepsilon_{0} \varepsilon\mathbf{E}$ .
The magnetization $\mathbf{M}$ is located in the thin plane layer
that is embedded into a dielectric surrounding with
permittivity $\varepsilon$ at a point
$z=0$. The magnetization can then be represented as%
\[
\mathbf{M}(t,\mathbf{r})=\mathbf{m}(t)n_{r}l\delta(z),
\]
where $l$\ is the film thickness, $n_{r}$\ is the density of SRR-JJ and
$\mathbf{m}(t)$ is the magnetic moment of the ring \cite{Maim:Gabi:10}.

\begin{figure}
\centerline{\epsfig{file=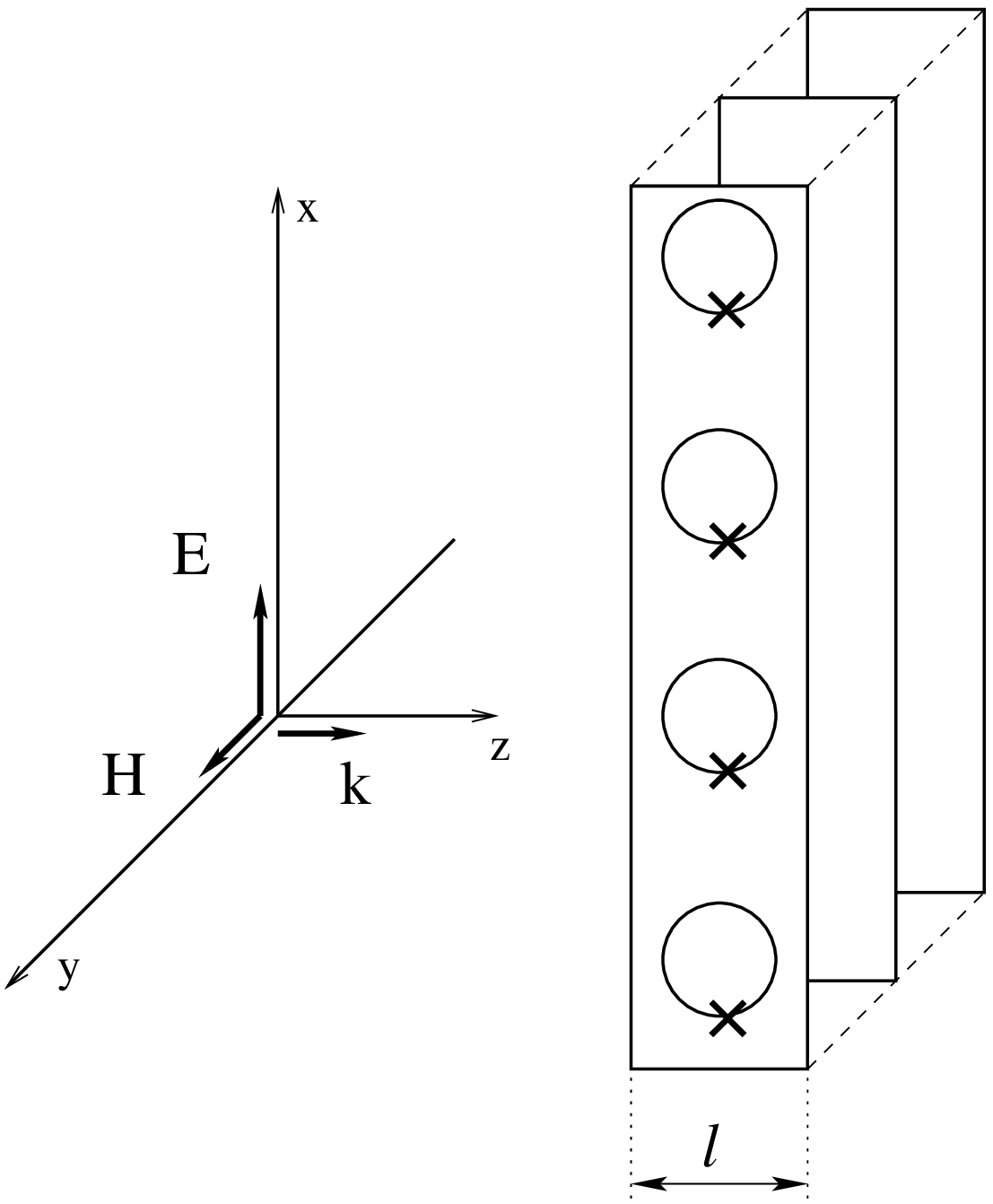,height=5cm,width=5cm,angle=0}
\epsfig{file=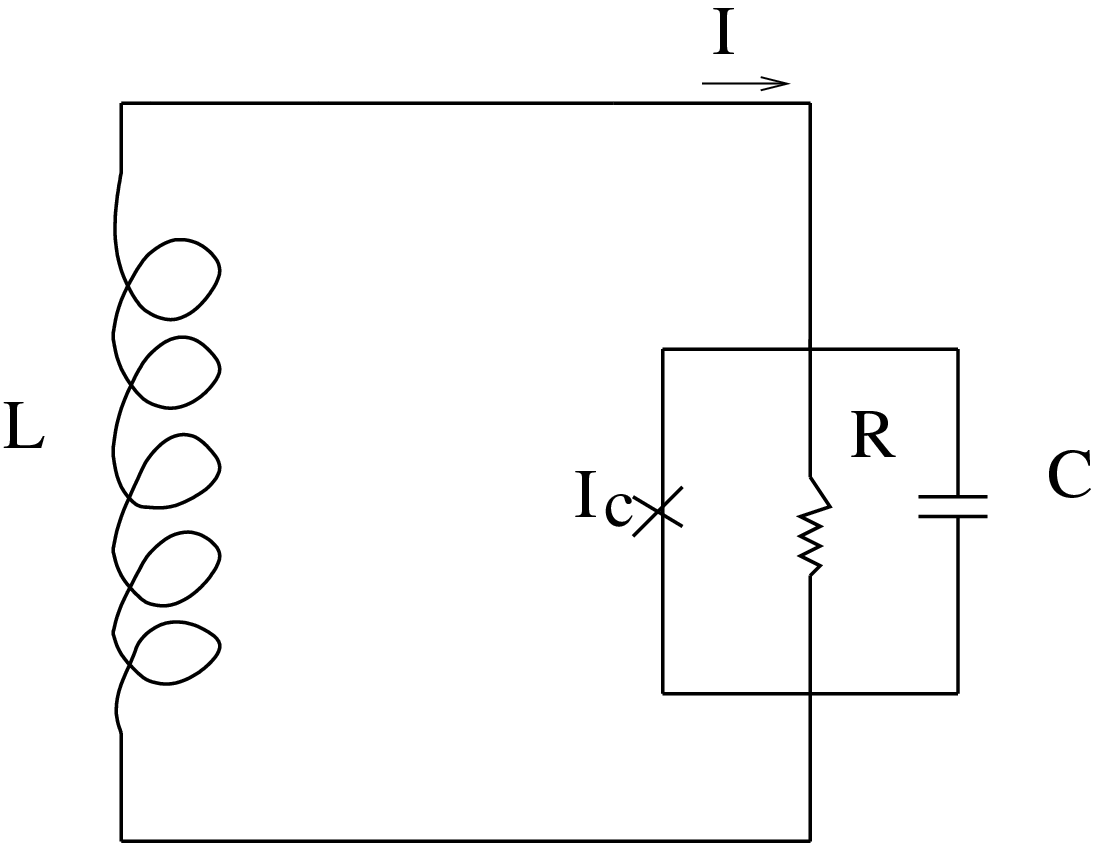,height=5cm,width=5cm,angle=0}}
\caption{The left panel shows a schematic view of the electromagnetic wave
incident on the split ring Josephson junction resonator in the
$(y,z)$ plane.
The right panel shows the equivalent circuit of the split-ring using
the Resistively Shunted Junction model for the Josephson junction. }
\label{f1}
\end{figure}

We assume that the plane electromagnetic wave propagates along the z-axis
and is polarized so that $\mathbf{H}$ is parallel to the y axis, the
normal to the plane of the SRR-JJ and $\mathbf{E}$ is parallel to the x
axis (see Fig. \ref{f1}). Since all the magnetic moments of the SRR-JJ in
the layer are parallel,  the magnetization
is parallel to $\mathbf{H}$. In
this case the Maxwell equations take the form
\[
E_{,z}=-(\mu_{0}H_{,t}+M_{,t}),\qquad
H_{,z}=-\varepsilon_{0}\varepsilon E_{,t}.
\]
They imply the wave equation
\begin{equation}
H_{,zz}-\mu_{0}\varepsilon_{0}\varepsilon H_{,tt}=\varepsilon_{0}
M_{,tt}\label{scalar_wave}
\end{equation}

Ohm's law applied to the ring gives
\[
\mathcal{E}=U_{L}+U_{C},
\]
where $\mathcal{E}$ is the electromotive force induced by the
magnetic field of the electromagnetic pulse incident on the ring.
\[
\mathcal{E}=-\frac{\partial\Phi}{\partial t}\approx- S H_{,t}(0,t),
\]
and where $ S $ is the surface enclosed by the ring. We neglect the
resistance of this loop which we assume to be made of
superconductive material. Following Josephson's second relation, the
voltage across the junction is
\[
U_{C}=\frac{\hbar}{2e}\frac{\partial\varphi}{\partial t}=\Phi_{0}
\frac{\partial\varphi}{\partial t},
\]
where $\varphi$ is the phase of the junction.
The inductance $L$ due to the loop of the SRR results in the voltage
\[
U_{L}=L\frac{\partial I}{\partial t}
\]
where $I$ is the current in the loop. Taking into account Ohm's law
together with these definitions we obtain
\[
\Phi+\Phi_{0}\varphi+LI=0,
\]
or
\begin{equation}
I=-L^{-1}(\Phi+\Phi_{0}\varphi).\label{current}
\end{equation}

The Resistive Shunted Junction model for the Josephson junction together
with Kirchoff's law for the current imply
$$
I=I_{C}+I_{J}+I_{R}=C\frac{\partial U_{C}}{\partial
t}+ I_c\sin\varphi
+ I_R$$
$$ =C\Phi _{0}\frac{\partial^{2}\varphi}{\partial
t^{2}}+I_c \sin\varphi
+ {\Phi _{0}\over R}\frac{\partial\varphi}{\partial t}
$$
Thus the variable $\varphi$ evolves according to
\[
C\Phi_{0}\frac{\partial^{2}\varphi}{\partial t^{2}}
+ {\Phi _{0}\over R}\frac{\partial\varphi}{\partial t}
+ I_c\sin\varphi
=-L^{-1}(\Phi+\Phi_{0}\varphi).
\]

In this equation the flux $\Phi$ has two components, the incident
magnetic flux and the flux induced by the current in the loop of the SRR-JJ.
That is
\[
\Phi=S H_{in}(t)-\frac{S}{2}\sqrt{\frac{\varepsilon_{0}}{\varepsilon\mu_{0}}
}\frac{\partial M}{\partial t}.
\]
Since $M=n_{r}l S I(t)$ we have
\[
M=-n_{r}l S L^{-1}(\Phi+\Phi_{0}\varphi).
\]

The wave equation (\ref{scalar_wave}) can be rewritten as
\[
H_{,zz}-\mu_{0}\varepsilon_{0}\varepsilon
H_{,tt}=\varepsilon_{0}\tilde {m}_{,tt}\delta(z)
\]
where the one ring magnetization $\tilde{m}$ can be defined as
\[
\tilde{m}(t)=n_{r}l S I(t).
\]
Taking into account the definition of the current (\ref{current}) one
can write
\[
\tilde{m}(t)=-n_{r}l S L^{-1}(\Phi+\Phi_{0}\varphi)=-n_{r}l S L^{-1}( S H+\Phi
_{0}\varphi).
\]
The phase variable $\varphi$ is governed by the equation
\[
\frac{\partial^{2}\varphi}{\partial t^{2}}
+ {1 \over R C } \frac {\partial\varphi}{\partial t} +\frac{I_c}
{C\Phi_{0}}\sin\varphi=-{1 \over L C } \frac{S H(0,t)}{\Phi_{0}} .
\]
The natural units of time, flux and space are given by
$$ \omega_T = 1/\sqrt{LC},~~  \Phi_0,
~~z_0 = {c \over \omega_T \sqrt{\varepsilon} },$$
where $ \omega_T$ is the
Thompson frequency and $z_0$ is the inverse of the Thompson wave number.
Hence, we can introduce the new dimensionless variables%
\[
h=S H/\Phi_{0},\qquad\tau=\omega_{T}t,\qquad\zeta=z/z_{0}.
\]
In terms of these variables the final equations are
\begin{equation}
h_{,\zeta\zeta}-h_{,\tau\tau}=-\kappa(h+\varphi)_{,\tau\tau}\delta
(\zeta),\label{wave}
\end{equation}
\begin{equation}
\varphi_{,\tau\tau}+\alpha \varphi_{,\tau}+ \varphi+\beta\sin\varphi
=-h,\label{srr_jj}
\end{equation}
where the parameters $\alpha,~~\beta$ and $\kappa$ are
$$ \alpha = {1\over R} \sqrt{L \over C},~~
\beta = {L I_c \over \Phi_{0} },~~
\kappa = \frac{n_{r}l S \varepsilon_{0}\omega_{T}}{L\sqrt{\varepsilon}}.$$

The system of equations (\ref{wave},\ref{srr_jj}) is our principal model
and we will investigate it in detail throughout the article.

Let us briefly recall the main properties of the SRR-JJ equation (\ref{srr_jj}).
It can be written as the 1st order system
\begin{eqnarray}
\varphi_{,\tau}  =\psi,\\
\psi_{,\tau}  =-\alpha \psi -\beta \sin(\varphi )-\varphi + h,
\end{eqnarray}
whose fixed points are $(0,0)$ and $(\varphi^*,0)$ where
\be\label{fix_relation} - \beta\sin(\varphi^* )- \varphi^* + h = 0.
\ee A plot of the above relation indicates that for  $\beta < 4.34$,
the equation has no solution so that there is only the $(0,0)$ fixed
point. For large $\beta$, the fixed points can be approximated using
an asymptotic expansion
$$\varphi^* = \varphi^{(0)} + {1 \over \beta} \varphi^{(1)} +
{1 \over \beta^2} \varphi^{(2)} + \dots$$ We get the approximation
of the fixed points as \be\label{approx_fix} \varphi^* = n \pi + {1
\over \beta} (-1)^n (h -n \pi) + \dots,\ee where $n$ is an integer.

\begin{figure}
\centerline{\epsfig{file=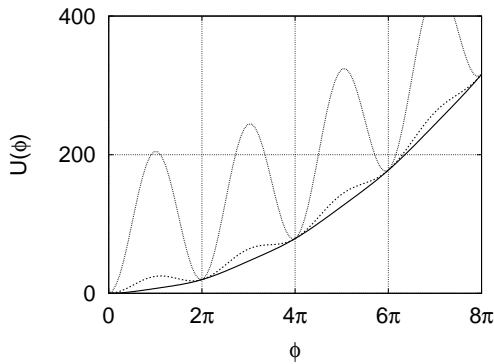,height=5cm,width=7 cm,angle=0}}
\caption{Potential energy $U(\varphi)= \beta (1-\cos\varphi) +
{1\over 2} \varphi^2$ with three different values of $\beta$, $\beta
= 1, 9.76 $ and $100$.} \label{f2}
\end{figure}

In the absence of damping $\alpha=0$ and forcing $h =0$, the system
is Hamiltonian with \be\label{ham} {\cal
H}(\varphi,\varphi_{,\tau})= {1\over 2} \varphi_{,\tau}^2 + \beta
(1-\cos\varphi) + {1\over 2} \varphi^2 \equiv {1\over 2}
\varphi_{,\tau}^2 + U(\varphi).\ee The stable fixed points
correspond to the minima of the potential $U(\varphi)$ and the even
values of the integer $n$. Fig. \ref{f2} shows a plot of the
potential $U(\varphi)$ for $\beta=1, 9.76$ and $100$. For $\beta=1$
shown as a continuous curve (red online) there is only one fixed
point $\varphi=0$. For $\beta=9.76$ shown in dashed line (green
online) there are three minima corresponding to stable fixed points,
$\varphi=0, \pm \phi^*$ where $\varphi^* \approx 2 \pi$. For
$\beta=100$ there are many stable fixed points.

\begin{figure}
\vskip-1.7cm \centerline{\epsfig{file=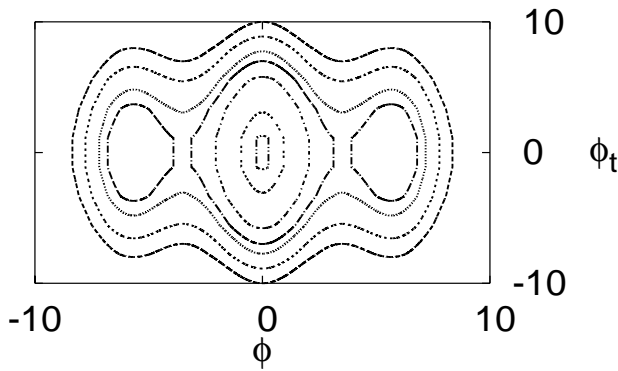,height=8 cm,width=12
cm,angle=0}} \vskip-1.7cm \caption{Phase portrait
$(\varphi,\varphi_\tau)$ of the Hamiltonian system (\ref{ham})
($\alpha=0$) for $\beta = 9.76$ and $\mathcal{H}=0$. The contour
levels presented are 0.1, 1, 5, 17., 25.0360499136927 (separatrix) ,
30.,40. and 50.} \label{f3}
\end{figure}

For this one degree of freedom Hamiltonian, the orbits are the
contour levels of the Hamiltonian. An important orbit is the
separatrix connecting the two unstable fixed points
$\varphi=\varphi^*\approx  \pi$. It is given by $ \mathcal{H} =
\mathcal{H}(\varphi^*,0)$.
%This value of the Hamiltonian can be approximated for $\beta >>1$ as
%\be\label{ham_separatrix}
%H(\phi^*,0) \approx {\pi^2 \over 2  }(1 + {2 \over \beta  })
%+ \beta(1+ \cos {\pi \over \beta})\approx {\pi^2 \over 2} + 2 \beta
%+{\pi^2 \over 2 \beta}.\ee
Fig. \ref{f3} shows the
phase portrait for $\beta = 9.76$.
For this value there are only
five fixed points. Notice the closed orbits around the
fixed points, the closed orbits surrounding the three stable fixed points.

\begin{figure}
\centerline{\epsfig{file=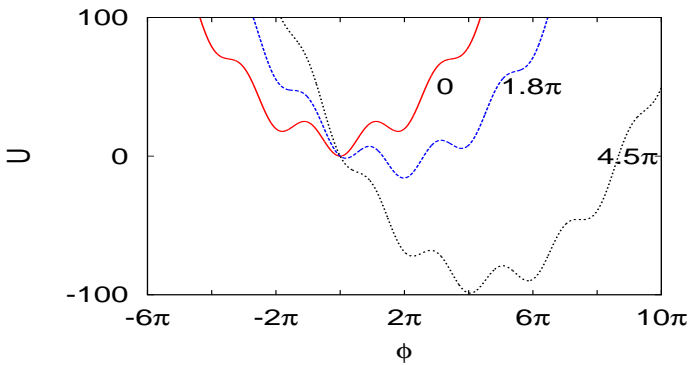,height=5cm,width=10 cm,angle=0}}
\caption{Potential energy $U(\varphi)= \beta (1-\cos\varphi) +
{1\over 2} \varphi^2 -h \varphi$ for three different values of the
static incident flux $h  = 0, 1.8 \pi$ and $4.5 \pi$.} \label{f2a}
\end{figure}

Another point is that the incident flux can be used to modify the
energy levels of the system. Assuming the incident flux to be
constant we can add a term to the potential and obtain the
generalized potential \be\label{pot2} U(\varphi)= \beta
(1-\cos\varphi) + {1\over 2} \varphi^2 - h \varphi,\ee where $h $ is
the normalized incident flux, assumed constant. This expression is
plotted in Fig. \ref{f2a} for $\beta=15$ and $h =0, 1.8 \pi$ and
$4.5 \pi$. The minima are symmetric for $h =0$ and they are shifted
to the left and the corresponding value of the potential is
decreased. By applying a sufficiently large continuous field one can
then shift the system from one state to the other. Notice that for
the large field $h=4.5 \pi$, there is a fourth minimum, around
$7\pi$.

To summarize, the SRR-JJ has energy states $U(\varphi)$ near
$2n\pi$. The number of these states is controlled by the parameter
$\beta$ i.e. the ratio of the flux to the flux quantum and the
magnetic field $h$. In the next section, we will select the incident
flux $h$ to move the system from one equilibrium to another.

%\section{Influence of damping }
\section{Switching between equilibrium states}

We consider now that the state of the system can be shifted from one
fixed point to another via an incident flux. For a short lived
perturbation, the system then relaxes freely to a minimum of energy.
The influence of the damping is essential , it should be present to
allow the relaxation but small to preserve the picture of the
potential. To examine how an incident flux will shift the system
from one equilibrium position to another it is useful to analyze the
work equation. To obtain it, we multiply  (\ref{srr_jj}) by
$\varphi_{,\tau}$ and integrate over time. We get the difference in
energy
\[
E(\tau_2)-E(\tau_1) \equiv \left[{1\over 2} \varphi_{,\tau}{}^2 +
\beta (1-\cos\varphi) + {1\over 2}
\varphi^2\right]_{\tau_1}^{\tau_2} =
\]
\be\label{work_eq} =\int_{\tau_1}^{\tau_2} d\tau h 
\varphi_\tau - \alpha \int_{\tau_1}^{\tau_2} d\tau \varphi_\tau^2 .
\ee The first term on the right hand side is the forcing while the
second one is the damping term. When a square pulse is applied to
the system, such that
$$h (\tau) = a ,~ {\rm for} ~ \tau_1 < \tau < \tau_2, ~~~0~~{\rm elsewhere}$$
the first integral is $a[\varphi(\tau_2)-\varphi(\tau_1)]$. If the
system is started at $(0,0)$ in phase space so that the
$E(\tau_1)=0$ and $\varphi(\tau_1)=0$, we have
$$E(\tau_2) = a\varphi(\tau_2) - \alpha \int_{\tau_1}^{\tau_2} d\tau  \varphi_\tau^2,$$
so that $\varphi(\tau_2)$ determines how much energy is fed into the
system. When the pulse is long $\tau_2 >> \tau_1$ $\varphi(\tau)$
will relax and oscillate so that there are values of $\tau_2$ such
that $\varphi(\tau_2) ~0$. In that case no energy gets fed into the
system. A sure way to avoid this is to take a narrow pulse.

The natural frequency of the oscillator around the $(0,0)$ fixed
point is \be\label{nat_freq} \omega_0 = \sqrt { \beta +1},\ee which
for $\beta =9.76$ gives $\omega_0\approx 3.28$ and a period $T_0 =
2\pi /\omega_0 \approx 1.91$. To simplify matters we now consider a
pulse of duration much smaller than $T_0$. This is experimentally
feasible. For the present analysis the pulse duration is very small
so that it can be approximated by the Dirac delta function
$h (\tau)= a \delta(\tau)$, where $a$ is a parameter. Let
us analyze briefly the solution. The equation (\ref{srr_jj}) becomes
\be\label{srr_jj_delta} \varphi_{\tau\tau} + (\alpha
+\delta)\varphi_{\tau} + \beta \sin(\varphi )+\varphi = a
\delta(\tau).\ee Integrating the equation on a small interval of
size $\epsilon$ around 0, we get \be\label{jump}
[\varphi_{\tau}]_{-\epsilon}^{\epsilon} + \alpha
[\varphi]_{-\epsilon}^{\epsilon} + \int_{-\epsilon}^{\epsilon} d\tau
(\beta \sin \varphi + \varphi) = a .\ee We now take the limit
$\epsilon \to 0$. We will assume continuity of the phase so that
$[\varphi]_{-\epsilon}^{\epsilon} \to 0$. The third term being the
integral of a continuous function tends to 0 when the bounds tend to
0. Assuming $\varphi_{\tau}(0_-)=0$ we get $\varphi_{\tau}(0_+)=a$
so that such a short incident pulse will just give momentum to the
system. With these initial conditions we solve equation
(\ref{srr_jj}) numerically using a Runge-Kutta algorithm with step
correction of order 4 and 5.

\begin{figure}
\centerline{\epsfig{file=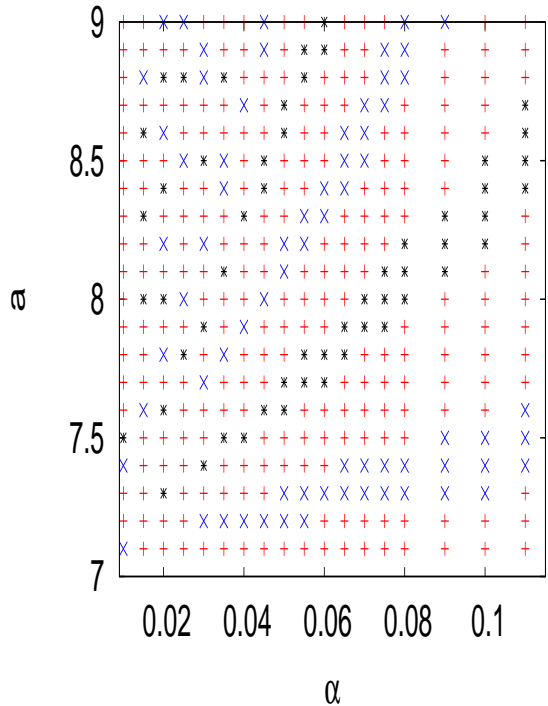,height=10  cm,width=8 cm,angle=0}}
\caption{Parameter plane $ (\alpha, a)$ showing the different
final states reached by the system, the left focus $*$ (blue online), the
center focus $+$ (red online) and the right focus $\times$ (green
online). The parameter $\beta = 9.76$.}
\label{f4}
\end{figure}

%\subsection{Analysis of the trapping: probabilistic approach }

We will now explore systematically the
plane $( \alpha, a)$ characteristic of the incident pulse.
The plot in the $ (\alpha, a)$ parameter plane shown in Fig. \ref{f4}
shows the final states, $O$ the central focus ($+$), $R$ the
right focus ($\times$) and $L$ the left focus $*$
reached by the system. Notice how these
are organized in "tongues" following the sequence $O~R~O~L~O~R~O~L \dots$
as one sweeps the plane counterclock-wise starting from the horizontal axis.
This simple geometrical picture can be understood by
examining Fig. \ref{f3}. In the case of small damping, the separatrices
around the fixed points are not affected very much.
Their perimeter is
proportional to the probability of reaching one fixed point or another.
The system is moving clock-wise along the orbits.
Assume the system reaches the central point $O$ for a given set of parameters.
If the damping is increased, the orbit might not reach $O$ but will
settle in $L$. Similarly if more kinetic energy is given to the
oscillator, it might reach $R$ instead of $O$.

Another important point is that the impulse given to the resonator must
be very short so that it relaxes following a free dynamics. The typical
frequencies of these devices are about 500 GHz so the impulse must be
around 5 Thz.

Now that we have seen how to prepare the system in a given state, it
is important to check if this state is really reached. For that one
can use a small signal to analyze the
state by reflection or transmission. This is the object of the next
section.

\section{Microwave spectroscopy of the SRR-JJ}

We assume that the ring is submitted to a fixed magnetic field $h_s$
and that it is in a local minimum $\varphi_s$ of the potential. Then
we send in a small electromagnetic pulse $\delta H$ and examine the
response $\delta \varphi$ of the ring using the scattering theory.
The linearized equations for $\delta h,~\delta \varphi$ read
\begin{eqnarray}
&& \delta h_{\zeta\zeta} -\delta h_{\tau\tau} = -\gamma \left (
-\delta  \Phi_{\tau\tau} + \delta h_{tt} \right )  \delta(\zeta)
,  \notag \\
[-1.5ex]  \label{dh_dphi} \\
&& \delta  \Phi_{\tau\tau} + \alpha \delta  \Phi_{\tau} + \beta
\cos(\varphi_s )\delta  \Phi +\delta  \Phi=\delta  h . \notag
\end{eqnarray}
We now assume periodic solutions \be\label{h_phi_harm}\delta h = h_a
e^{i \omega \tau},~\delta \varphi =\varphi_a e^{i \omega \tau},\ee
and obtain the reduced system
\begin{eqnarray}
&& h_{a,\zeta\zeta} +\omega^2 h_a = \gamma \omega^2 \left (
-\varphi_a + h_a \right )  \delta(\zeta)
,  \notag \\
[-1.5ex]  \label{h_phi_z} \\
&& \left [-\omega^2 + i \alpha \omega +1 + \beta \cos(\varphi_s
)\right ]\varphi_a =h_a . \notag
\end{eqnarray}
In the scattering we assume the electromagnetic wave to be incident
from the left of the film located at $\zeta=0$. We then have
\be\label{scat} h_a =  e^{-i \omega \zeta} + \mathcal{R} e^{i\omega
\zeta} ~,\zeta<0~;~~~~ h_a = \mathcal{T}e^{-i \omega \zeta }, \zeta
>0~~,\ee where $\mathcal{R}$ is the amplitude of the reflected wave
and $\mathcal{T}$ the amplitude of the transmitted wave. We have the
following interface conditions at $\zeta=0$ \be\label{interface}
h_a(0^-)= h_a(0^+),~~[h_{a,\zeta}]_{0^-}^{0^+} = \omega^2 \gamma
\left ( -\varphi_a(0) + h_a(0) \right ).\ee They imply two equations
for $\mathcal{R}$ and $\mathcal{T}$
%{\small \begin{eqnarray} 1+R & =& T, \notag \\
%-T -(-1+R) & =& -i \omega \gamma T
%\left [ {-1 \over -\omega^2 + i \alpha \omega +1 + \beta
%\cos(\phi_s ) } +1 \right ] ,\notag  \end{eqnarray} }
from which we obtain the transmission coefficient
\be\label{transmission} \mathcal{T} = { 2(-\omega^2 + 1 +\beta \cos
\varphi_s + i \omega \alpha ) \over D},\ee the reflection
coefficient \be\label{trans} \mathcal{R} = { -\alpha \gamma\omega^2
+ i\gamma\omega (\beta \cos \varphi_s - \omega^2) \over D},\ee and
where the denominator is \be\label{den_trans_ref} D = 2(-\omega^2 +
1+ \beta \cos \varphi_s ) + \alpha \gamma\omega^2 + i [2\alpha\omega
- \gamma\omega(\beta \cos \varphi_s-\omega^2)] .\ee

The square of the modulus of $R$ is \be\label{ref2} |\mathcal{R}|^2
= { \alpha^2 \gamma^2\omega^4+ \gamma^2\omega^2(\beta \cos \varphi_s
- \omega^2)^2 \over |D|^2} .\ee As seen in section 3, the split-ring
oscillator has a finite number of equilibria $\varphi_s$ depending
on the parameter $\beta$. We will assume that $\alpha=0$ throughout
the section. This damping does not change the width of the
resonances, it only affects the bounds of $|R|$. For $\alpha=0,~
0<|R|<1$ while for $\alpha> 0,~ \alpha <|R|<1-\alpha$. When
$\alpha=0$ $|\mathcal{R}|^2$ is an even function of $\omega$,
therefore we will only consider $\omega >0$.

As a first example we consider $h=0$ and $\beta=9.76$ corresponding
to the potential plotted in Fig. \ref{f2}.
\begin{figure}
\centerline{ \epsfig{file=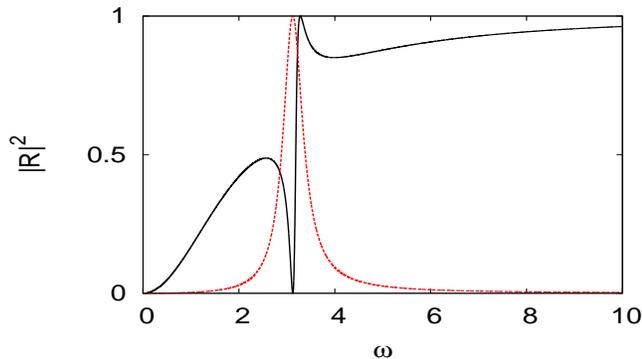,height=5  cm,width=9  cm,angle=0}
} \caption{Square of the modulus of the reflection coefficient
$|R|^2$ as a function of the frequency $\omega$ for
$\beta=9.76,~h=0,~\varphi_s=0$ for a large coupling $\gamma=1$. The
dashed line (red online) presents the reflection coefficient $|R|^2$
for a polarized dielectric film for comparison.} \label{f5d}
\end{figure}
The reflection coefficient is plotted in Fig. \ref{f5d} for
$\gamma=1$ and $\varphi_s=0$. Notice how the resonance is asymetric
around a minimum $\omega_s$. The system is passing for low
frequencies only. This line shape is a typical Fano resonance
\cite{nm11} due to the nonlocal coupling associated to the
$\varphi_{\tau\tau}$ term in the right hand side of the wave
equation (\ref{wave}). To see this, we have reported in Fig.
\ref{f5d} the reflection coeffcient for a simple polarized
dielectric film \cite{ckm06}. This simpler system is passing for all
frequencies except for a narrow band around the resonant frequency
$\omega_s$. The expression for the resonant frequencies $\omega_s$
can be obtained by considering the minima of $|\mathcal{R}|^2$.
These correspond to the second term in the numerator of (\ref{ref2})
being zero. We get \be\label{zeroesofr} \omega_s =
\sqrt{\beta\cos\varphi_s} .\ee

\begin{figure}
\centerline{
\epsfig{file=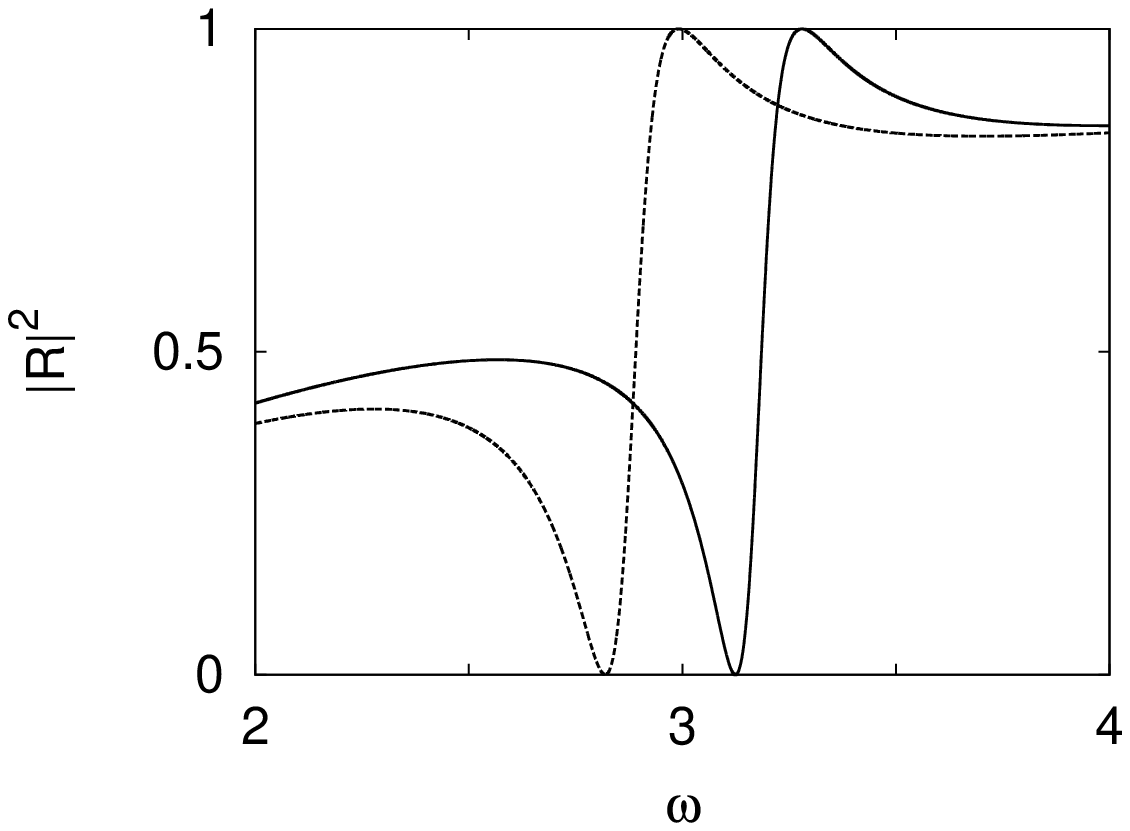,height=5  cm,width=4 cm,angle=0}
\epsfig{file=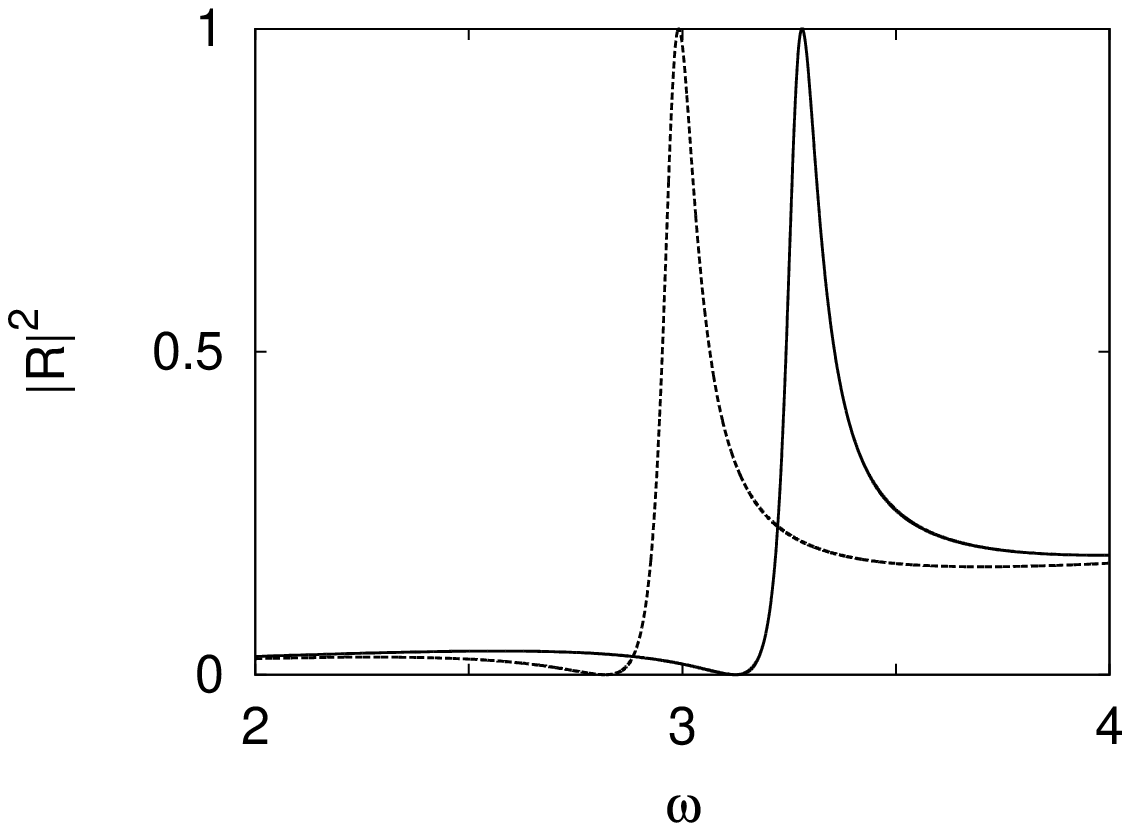,height=5  cm,width=4 cm,angle=0}
}
\caption{Square of the modulus of the reflection coefficient $|R|^2$
as a function of the frequency $\omega$ for $\beta=9.76$ and $H=0$
for a large coupling $\gamma=1$ (left panel) and a small coupling
$\gamma=0.2$ (right panel). }
\label{f5a}
\end{figure}

Let us now examine the influence of the different parameters on the
reflection coefficient. Fig. \ref{f5a} shows the reflection
coefficient $|R|^2$ for the two steady states $\varphi_s$ and
$\gamma=1$ (left panel) and $\gamma=0.2$ (right panel).
For the large coupling the resonance is sharp although for large
$\omega$ we recover $|R|=1$. The value of the steady state
$\varphi_s$ also influences  $|R|$. In Fig. \ref{f5a}, the
continuous curve corresponds to $\varphi_s=0$ and the dashed line
corresponds to $\varphi_s= \pm 5.66$, the two additional local
minima. The two curves are very close and the reflection coefficient
does not indicate if we are in the left or right minimum.  Notice
also that the system only passes low frequencies for $\gamma=1$. On
the contrary for $\gamma=0.2$ it is globally passing for all
frequencies except for a narrow band around $\omega_s$. These plots
would enable to extract the coupling parameter from a SRR-JJ.

\begin{figure}
\centerline{\epsfig{file=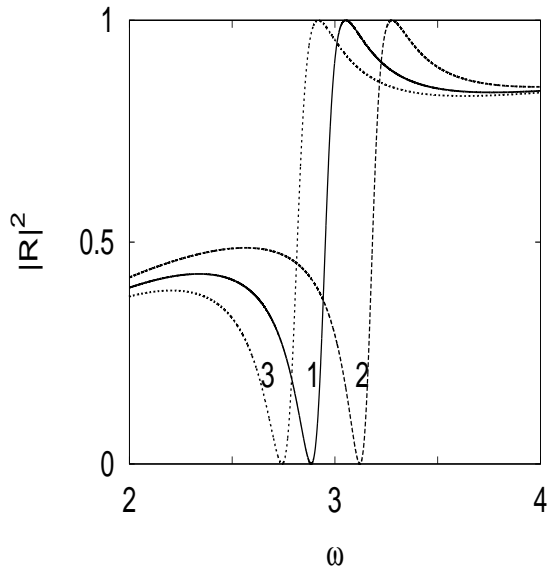,height=8  cm,width=8 cm,angle=0}}
\caption{Square of the modulus of the reflection coefficient $|R|^2$
as a function of the frequency $\omega$ for the three different
equilibria indicated in Table I. The parameters are $\beta=9.76,~~h
=1.8\pi,~~\gamma=1$ and $\alpha=0$.} \label{f5c}
\end{figure}

As a second example we consider $\beta=9.76$ and $h=1.8\pi$
corresponding to the potential plotted in Fig. \ref{f2a}. There are
three minima given in the table I. The reflection coefficient
$|R|^2$ is plotted in Fig. \ref{f5c}. Now it is possible to
distinguish the "left" minimum 1 from the "right" minimum 3 because
the non zero magnetic field $h$ has lifted the degeneracy.
\begin{table} \label{tab2}
\begin{tabular}
{| c | c | c | c |}
  \hline
${\rm index }$   &  1    &  2 &  3 \\ \hline $\varphi_s$   &  0.550
& 6.224 &  11.875\\ \hline $\omega_s$ &  2.884       &  3.121   &
2.742  \\\hline
 \end{tabular}
\caption{Spectroscopy data $(\omega_s,\varphi_s)$ for a split-ring
oscillator with three steady states. The parameters are
$\beta=9.76,~ ~h =1.8\pi,~\gamma=1,~\alpha =0$.}
\end{table}

\begin{figure}
\centerline{\epsfig{file=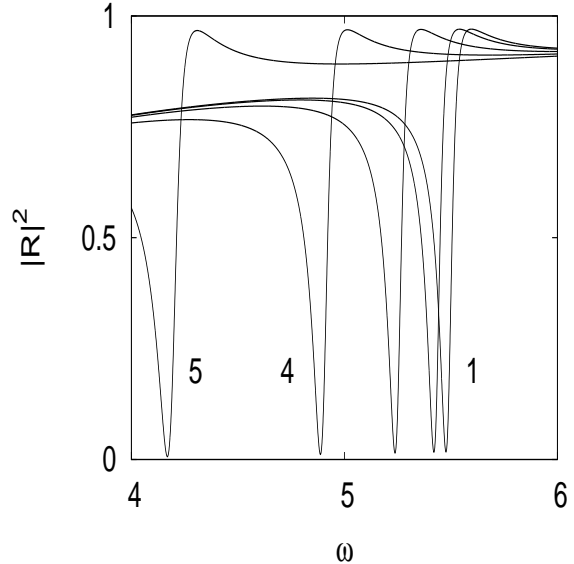,height=8  cm,width=8 cm,angle=0}}
\caption{Square of the modulus of the reflection coefficient $|R|^2$
as a function of the frequency $\omega$ for the five different equilibria
indicated in Table II. The index of the minima is indicated in the
plot. The parameters are $\beta=30,~~\gamma=1$
and $\alpha=0.01$.}
\label{f5b}
\end{figure}

Finally we consider $h=0$ and $\beta=30$ for which the potential
$U(\phi)$ has five minima indicated in the table II.  The square of
the modulus of the reflection coefficient (\ref{ref2}) is plotted in
Fig. \ref{f5b} for the five different equilibria. In the example
shown, the spectroscopy data $(\omega_s,\varphi_s)$ is given in
table II. Again there is no possibility to distinguish whether the
system is in the left or the right minimum.
\begin{table} \label{tab1}
\begin{tabular}
{| c | c | c | c | c | c |}
  \hline
${\rm index }$   &  1    &  2 &  3&  4 & 5 \\ \hline $\varphi_s$   &
0    &  6.08 &  12.15&  18.2 & 24.18 \\ \hline $\omega_s$ & 5.477 &
5.420 & 5.238 & 4.888 & 4.169   \\\hline
 \end{tabular}
\caption{Spectroscopy data $(\omega_s,\varphi_s)$ for a split-ring
oscillator with five steady states. The parameters are  $\beta=30,~
\gamma=1,~\alpha =0.01$.}
\end{table}

\section{Conclusion }

We considered the interaction of an electromagnetic wave with
a split-ring Josephson resonator. For appropriately chosen
parameters, the resonator has excited states whose number can be
controlled by carefully tuning the inductance and capacity of the
ring. The existence of these excited states makes this system similar
to an artificial atom with discrete energy levels. Other artificial atoms
containing nonlinear elements like a diode \cite{Lapin:2003}, a Kerr
material \cite{Zharov:2003} or a laser amplifier \cite{Gabitov_Kennedy:2010}
would not give these discrete levels. In addition, since
the oscillator is operating in the superconducting regime the losses
are very small as opposed to the current metamaterials.

First we assumed that there are just two excited states and showed
how an incident magnetic flux can shift the system from the ground
state to one of these excited states. By sending a microwave field
on the resonator we can perform a spectroscopy of it and characterize
in which state it is. Using a scattering theory formalism we computed
the reflection and transmission coefficients for the wave. These have
a specific minimum that we can compute analytically. Another interesting
feature is the form of the resonance which is Fano like and very
different from the one obtained for a ferroelectric thin
film \cite{ckm06}. Note that
the right and left minima are undistinguishable with this
measurement and that a static magnetic field will lift this degeneracy.
We hope this analysis will be useful to experimentalists studying
split ring Josephson resonators.

{\bf Acknowledgements} \\
The authors thank Matteo Cirillo, Alexei Ustinov and Alexey Yulin for very helpful
discussions. JGC and AM thank the University of Arizona for
its support. AIM is grateful to the
Laboratoire de Math\'ematiques, INSA de Rouen for hospitality and
support. The computations were done at the Centre de Ressources
Informatiques de Haute-Normandie. This research was supported by RFBR grants No.
09-02-00701-a.

\end{document}